\documentclass[conference]{IEEEtran}
\IEEEoverridecommandlockouts
\usepackage{cite}
\usepackage{amsmath,amssymb,amsfonts}
\usepackage{algorithmic}
\usepackage{graphicx}
\usepackage{mwe}
\usepackage{textcomp}
\usepackage{dsfont}
\usepackage{booktabs}
\usepackage{xcolor}
\def\BibTeX{{\rm B\kern-.05em{\sc i\kern-.025em b}\kern-.08em
    T\kern-.1667em\lower.7ex\hbox{E}\kern-.125emX}}

\begin{document}

\title{Latent Semantic State Estimation for Reliable Swarming of UAVs under Intermittent Connectivity\\
\thanks{This research was supported by the U.S. National Science Foundation under Grant CNS-2210254.}
\thanks{\copyright~2026 IEEE. Personal use of this material is permitted.
Permission from IEEE must be obtained for all other uses, in any current or future media, including reprinting/republishing this material for advertising or promotional purposes, creating new collective works, for resale or redistribution to servers or lists, or reuse of any copyrighted component of this work in other works.}
}

\author{\IEEEauthorblockN{Paris A. Karakasis and Walid Saad}
\IEEEauthorblockA{\textit{Institute for Advanced Computing} and \textit{Bradley Dept. of Electrical and Computer Engineering}\\
\textit{Virginia Tech}, Alexandria, VA, USA \\
Emails: \{parisk, walids\}@vt.edu}}

\maketitle

\begin{abstract}

Cooperative multi-unmanned aerial vehicle (UAV) reconnaissance is often hindered by intermittent air-to-air communications where link dropouts lead to uncoordinated exploration and redundant mapping. Existing approaches rely on explicit exchange of high-dimensional spatial data or raw observations, incurring significant overhead, and often revert to reactive individual exploration during outages. This paper proposes a memory-augmented framework in which each UAV maintains a structured latent state decomposed into map, task, and memory components. During dropout, a generative predictor conditioned on the memory state infers substitute peer messages in the latent space, making the estimation task more tractable and directly aligned with the cooperative objective. The framework is trained end-to-end under the centralized training with decentralized execution paradigm. Simulation results demonstrate that the proposed framework closely matches the performance of a fully connected swarm, while remaining robust across a wide range of link failure conditions.

\end{abstract}

\begin{IEEEkeywords}
UAV swarms, cooperative reconnaissance, intermittent communication, latent state estimation, multi-agent reinforcement learning, memory-augmented coordination
\end{IEEEkeywords}


\section{Introduction}
\label{sec:intro}

Unmanned aerial vehicle (UAV) swarms are increasingly deployed in safety-critical applications such as search-and-rescue, environmental monitoring, and surveillance~\cite{puri2005uav}. In these settings, inter-agent coordination and regular exchange of observations and task-relevant information is essential for effective UAV task completion~\cite{campion2018swarm}. In cooperative reconnaissance tasks, UAVs naturally tend to disperse across the environment to cover it efficiently. The resulting dispersion increases the likelihood of air-to-air (A2A) wireless communication link degradation. Unlike tightly coupled swarms operating in proximity, the UAVs here form a loosely coordinated \emph{swarm} where spatial dispersion is a feature of effective operation rather than an anomaly. As a result, coordination under intermittent connectivity emerges as a core design requirement rather than an edge case.

In real UAV deployments, A2A communication links are inherently unreliable. Factors such as distance-dependent path loss, multipath fading at low altitudes, beam misalignment from rapid maneuvers, co-channel interference, atmospheric attenuation, energy constraints, and hardware failures can all cause intermittent link dropout~\cite{yan_survey, goddemeier2015, matolak_survey, gupta2015survey, mozaffari2019tutorial, itu_p838}. 
Hence, communication dropout is an expected feature of real deployments, directly impairing coordination and mission performance. This poses a fundamental question: \emph{How can UAVs remain effectively coordinated when communication with the swarm is temporarily unavailable?}
Answering this question requires overcoming three challenges. First, the UAV swarm must maintain safe separation between its members, which requires each UAV to know all its neighbors' states. Second, without peer information, UAVs may revisit already-covered areas which can directly degrade collective performance. Third, predicting raw observations of all collaborating UAVs in the swarm becomes computationally prohibitive as the swarm grows. Thus, there is a need for compact representations that can capture, out of all the observations and states of the other UAVs, only the necessary information needed for coordination.

\subsection{Related Works}
\label{sec:related}


Prior works have addressed partial aspects of these challenges~\cite{semcom_survey_qin, semcom_survey_xie, commnet, dial, message_dropout, fcmnet, zhang2024macsl, aicoral2025, santos2025, mahdoui2020, westheider2023, zhang2024}. Semantic communication~\cite{semcom_survey_qin} replaces the classical goal of bit-level reconstruction with preserving the meaning of the source data at the receiver, while task-oriented communication is narrower still, retaining only what the receiver needs for its downstream task. In~\cite{semcom_survey_xie}, an encoder-decoder pair is trained end-to-end to learn a low-dimensional representation sufficient for the receiver's purpose. In~\cite{commnet} and \cite{dial}, the idea is extended to multi-agent reinforcement learning (MARL) to discover effective messaging strategies. However, these works assume reliable wireless links and do not address communication dropout. To improve robustness, the authors in~\cite{message_dropout} and \cite{fcmnet} employ regularized learning, but they treat communication dropouts as noise to be tolerated rather than information to be recovered. More recently, the authors in~\cite{zhang2024macsl}, \cite{aicoral2025}, and \cite{santos2025} propose using generative models to reconstruct missing information via recurrent networks, variational auto-encoders, and auto-regressive imputation, respectively. However, all three operate on raw observations or generic, non-task-oriented features, making reconstruction costly and only loosely aligned with the coordination objective. A shared limitation of all the approaches above is that none of them learns a compact, task-oriented latent representation capable of sustaining coordination under intermittent communication.

Recent works in multi-UAV cooperative mapping and exploration, such as \cite{mahdoui2020} and \cite{westheider2023}, explicitly communicate spatial data, e.g., frontier points, poses, and occupancy grid updates. However, these solutions incur significant communication overhead and lack any mechanism to maintain coordination during outages. 
While some works have introduced memory-based representations for individual exploration guidance \cite{zhang2024}, these representations are not directly designed to capture information related to the other UAVs in the swarm. Hence, they cannot be reliably used to substitute missing peer information.


\subsection{Contributions}

The main contribution of this paper is a memory-augmented framework for cooperative UAV reconnaissance that maintains effective coordination under intermittent A2A communication. Our approach enables each UAV, within a swarm, to maintain a coherent internal representation that supports both local decision-making and swarm coordination, even when links to the other UAVs in the swarm are unavailable. To this end, we make the following contributions:
\begin{itemize}
    \item We design a structured latent state, maintained locally by each UAV and decomposed into map, task, and memory components, that enables task-oriented communication focused solely on coordination.
    \item We introduce a memory-based generative predictor that locally infers the missing aggregated message based on each UAV's accumulated latent history.
    \item We propose an asymmetric update strategy that shields map estimates from hallucinated updates during dropout.
\end{itemize}

Unlike the approaches covered in Section~\ref{sec:related}, our method operates on compact task-oriented and semantic representations, \emph{defined as latent encodings retaining only the coordination-relevant content of each UAV's observations, and as map compressions decoding to the swarm's collective map estimate, respectively}. This design choice makes prediction of the missing aggregated peer message more tractable and directly aligned with the cooperative objective. The proposed framework is trained end-to-end under the centralized training with decentralized execution (CTDE) paradigm~\cite{ctde_intro}, and at execution time each UAV operates fully autonomously using only local observations and available messages. Simulation results show that the proposed approach closely matches the performance of a fully connected swarm across a wide range of link failures. This indicates that intermittent communication need not impose a severe performance penalty, as the swarm learns to coordinate nearly as well as if all links were reliable.


\section{System Model and Task Formulation}
\label{sec:system}


We consider a set $\mathcal{N}$ of $N$ UAVs forming a cooperative swarm tasked with exploring and mapping an unknown environment. 
Each UAV operates based on local observations and limited communication with its collaborative peers, making coordination under partial information essential. The collective goal is to build an accurate and complete map while maintaining effective swarm coordination during exploration. 

The cooperative reconnaissance task can be formalized as a decentralized partially observable Markov decision process (Dec-POMDP), defined by the tuple $\mathcal{M} = \langle \mathcal{N}, \mathcal{S}, \{\mathcal{A}^i\}, \mathcal{T}, \mathcal{R}, \mathcal{O}, P_o, \gamma \rangle$. Given a discretized spatial grid $\mathcal{X}$, the ground truth occupancy map $\boldsymbol{M}:\mathcal{X} \to \{0,1\}$ indicates which areas are free (0) or occupied (1). Through exploration, each UAV incrementally contributes to a collectively constructed estimate $\hat{\boldsymbol{M}}_t:\mathcal{X} \to \{0,1,-1\}$, where $-1$ is used for cells that have not yet been observed up to time $t$. Hence, the environment at time $t$ is described by a global state $\boldsymbol{s}_t = (\{\boldsymbol{e}_t^i\}_{i \in \mathcal{N}}, \hat{\boldsymbol{M}}_t)\in\mathcal{S}$. Here, $\boldsymbol{e}_t^i = [\boldsymbol{p}_t^i;\, \boldsymbol{v}_t^i;\, \chi_t^i;\, b_t^i]$ is the proprioceptive state of UAV $i$, comprising its position $\boldsymbol{p}_t^i \in \mathbb{R}^3$, velocity $\boldsymbol{v}_t^i \in \mathbb{R}^3$, heading $\chi_t^i\in\mathbb{R}$, and remaining energy level $b_t^i\in\mathbb{R}_+$. In our model, UAVs move in continuous 3D space. The grid only discretizes the environment for mapping and coverage. The energy of each UAV depletes at a constant rate $\kappa$ per time step from an initial budget of $b_0^i=100$. The complete reconstruction of $\boldsymbol{M}$, i.e., $\hat{\boldsymbol{M}}_{t^*}$, is nonnegative at some time $t^*$ and constitutes the task completion criterion. The mission terminates early if any UAV depletes its energy.

The global state $\boldsymbol{s}_t$ is not directly accessible to any single UAV. Instead, UAVs rely on a local observation $\boldsymbol{o}_t^i \in \mathcal{O}$ generated stochastically as $\boldsymbol{o}_t^i \sim P_o(\boldsymbol{o}_t^i \mid \boldsymbol{s}_t, \boldsymbol{p}_t^i)$, reflecting the limited and noisy nature of onboard sensing. When available, communication with peers is also used when a UAV selects an action $\boldsymbol{a}_t^i \in \mathcal{A}^i$. The action space $\mathcal{A}^i$ consists of feasible velocity commands subject to operational constraints. The joint action $\boldsymbol{a}_t = [\boldsymbol{a}_t^1; \dots; \boldsymbol{a}_t^N]$ drives the state transition $\boldsymbol{s}_{t+1} \sim \mathcal{T}(\boldsymbol{s}_{t+1} \mid \boldsymbol{s}_t, \boldsymbol{a}_t)$, where $\mathcal{T}$ is a stochastic transition kernel governing UAV motion dynamics. Following the cooperative MARL paradigm, all UAVs share a common reward $r_t = \mathcal{R}(\boldsymbol{s}_t, \boldsymbol{a}_t)$, defined in Section~\ref{sec:reward}, and cooperate to maximize the expected discounted return $\mathbb{E}\!\left[\sum_{t=0}^{T} \gamma^t r_t\right]$, where $\gamma \in [0,1)$ is a discount factor.

The cooperative nature of the task fundamentally distinguishes a UAV swarm from a set of independent ones. Since each UAV observes only a limited portion of the environment,
$\boldsymbol{M}$ cannot be reconstructed by a single UAV acting alone, and
effective coordination requires each UAV to maintain awareness of its peers'
progress. When links are unavailable, each UAV must rely on its current local
observation and accumulated history (past observations and communications) to approximate this collective awareness. The partial observability of $\boldsymbol{s}_t$ makes explicit state tracking intractable. This motivates the use of learned latent representations 
that compress partial, noisy observations into a compact, task-aligned summary that is cheap to exchange and can be tractably inferred from history when links drop. We call such a representation \emph{semantic} when it admits a decoder to an external referent under an explicit distortion measure, so that its content is fixed by what it represents rather than only by the action it induces. Section~\ref{sec:latent} makes this concrete for the map component of the exchanged message.


\subsection{Intermittent Communication Model}
\label{sec:comm}

Inter-UAV communication can be modeled as a time-varying directed graph
$\mathcal{G}_t = (\mathcal{N}, \mathcal{E}_t)$, where a directed edge
$(j, i) \in \mathcal{E}_t$ exists if and only if UAV $i$ successfully receives
a transmission from UAV $j$ at time $t$. Hence, the set of active neighbors of UAV $i$
at time $t$ is defined as $\mathcal{N}_t^i = \{j \in \mathcal{N} :
(j,i) \in \mathcal{E}_t\}$. The topology of $\mathcal{G}_t$ is governed by the underlying A2A wireless channel. For UAV-to-UAV links, a Rician fading model is appropriate, as A2A propagation typically features a dominant line-of-sight (LoS) component alongside scattered multipath contributions~\cite{yan_survey, goddemeier2015}, and \cite{matolak_survey}. 

The received signal-to-interference-plus-noise ratio (SINR) at UAV $i$ from UAV $j$ at time $t$ is given by:
\begin{equation}
\small
    \Gamma_{t}^{j\rightarrow i} = \frac{P^j_t  }
    {\sigma^2 + I_t^i} \cdot |h_{t}^{j\rightarrow i}|^2 \cdot \|\boldsymbol{p}_t^i - \boldsymbol{p}_t^j\|^{-\alpha} \cdot L^\mathrm{atm}_t \cdot \xi_{t}^{j\rightarrow i},
    \label{SINR}
\end{equation}
where $P^j_t$ is the transmit power of UAV $j$ at time $t$, $\sigma^2$ is the noise
power assumed fixed and common across all UAVs, $I_t^i$ is the time-varying
co-channel interference~\cite{zhou2021interference, khuwaja2019coci}, $\alpha$ is the path loss exponent capturing degradation as UAVs disperse, $L^\mathrm{atm}_t\in(0,1]$ is an atmospheric attenuation factor~\cite{itu_p838}, $h_{t}^{j\rightarrow i}$ is the
complex Rician fading coefficient modeling small-scale A2A fading, whose
$K$-factor grows with altitude~\cite{goddemeier2015, moraitis2023survey}, and
$\xi_{t}^{j\rightarrow i} \in \{0,1\}$ is a Bernoulli indicator of abrupt link
failure. A link $(j,i)$ is considered active at time $t$ if
$\Gamma_{t}^{j\rightarrow i} \geq \beta_{\mathrm{th}}$.

\subsection{Cooperative Reward Formulation}
\label{sec:reward}

Evaluating collective performance requires access to information that no single UAV can observe locally. This information is captured by the full global state $\boldsymbol{s}_t$, which defines a shared reward signal $r_t = \mathcal{R}(\boldsymbol{s}_t, \boldsymbol{a}_t)$ measuring the swarm's joint progress toward the reconnaissance objective. 

Our reward consists of three parts
$r_t = w_{\text{cov}}\, r_t^{\text{cov}} 
          + w_{\text{red}}\, r_t^{\text{red}} + w_{\text{coll}}\, r_t^{\text{coll}}$,
where $w_{\text{cov}}, w_{\text{red}}, w_{\text{coll}} > 0$ are weighting coefficients. 
The reward $r_t^{\text{cov}}$ incentivizes efficient exploration of new areas. Let $C(\hat{\boldsymbol{M}}_t)$ be the set of cells observed at least once by any UAV up to time $t$. The marginal increase in collective coverage, penalized by a constant $c > 0$ to discourage stalling, yields
$r_t^{\text{cov}} = |C(\hat{\boldsymbol{M}}_t)| - |C(\hat{\boldsymbol{M}}_{t-1})| - c$.
This term is computed centrally during training, as it requires $\hat{\boldsymbol{M}}_t$ over consecutive steps. The penalty $r_t^{\text{red}}$ discourages UAVs from revisiting areas already covered by their peers. Let $C^i_t \subseteq C(\hat{\boldsymbol{M}}_t) \setminus C(\hat{\boldsymbol{M}}_{t-1})$ be the set of cells newly observed by UAV $i$ at time $t$. Then, $r_t^{\text{red}} = -\sum_{i < j} \bigl|C^i_t \cap C^j_t\bigr|$, i.e., the pairwise explored region overlap. The penalty $r_t^{\text{coll}}$ discourages unsafe proximity, penalizing UAV pairs closer than $d_{\text{safe}}$, i.e., $r_t^{\text{coll}} = -\sum_{i < j} \mathds{1}\!\left[\|\boldsymbol{p}_t^i - \boldsymbol{p}_t^j\| < d_{\text{safe}}\right]$. 

Next, we must solve the core problem of sustaining coordinated, efficient, and collision-free exploration when the messages of the other UAVs in the swarm fail to arrive. 


\section{Memory-Augmented Coordination under Communication Dropout}
\label{sec:architecture}


Our framework is trained end-to-end under the CTDE paradigm and is built around a structured latent state decomposition, an asymmetric dropout strategy that preserves map integrity, and a memory-based predictor inferring missing messages from the UAV's accumulated latent history.


\subsection{Latent Representation and Memory}
\label{sec:latent}

We decompose the latent state of UAV $i$ into three components. The \emph{map state} $\boldsymbol{\mu}_t^i \in \mathbb{R}^{d_{\mu}}$ encodes a compressed estimate of the occupancy map $\hat{\boldsymbol{M}}_t$. The \emph{task state} $\boldsymbol{\tau}_t^i \in \mathbb{R}^{d_{\tau}}$ encodes the UAV's operational status and swarm coordination state, driving decentralized decisions. The \emph{memory state} $\boldsymbol{\rho}_t^i \in \mathbb{R}^{d_{\rho}}$ recurrently accumulates a compressed history of observations, task states, and past communications, providing the context for predicting missing aggregated messages during outages.

\subsubsection{Map state update} The map state is updated via a recurrent function $\boldsymbol{E}_{\boldsymbol{\phi}}$. Given the aggregated map message $\boldsymbol{m}_t^{\mu,i}$ (Section~\ref{sec:comms}), previous $\boldsymbol{\mu}_{t-1}^i$ gets updated with the UAV's current local observation and position by $\boldsymbol{\mu}_t^i = \boldsymbol{E}_{\boldsymbol{\phi}}\!\left(\boldsymbol{\mu}_{t-1}^i,\; \boldsymbol{o}_t^i,\; \boldsymbol{p}_t^i,\; \boldsymbol{m}_t^{\mu,i}\right)$.
During training, $\boldsymbol{E}_{\boldsymbol{\phi}}$ and a decoder
$\boldsymbol{D}_{\boldsymbol{\eta}}$ are trained jointly under a map
reconstruction objective (Section~\ref{sec:training}), ensuring that
$\boldsymbol{\mu}_t^i$ serves as a compressed approximation of the swarm's current map estimate $\hat{\boldsymbol{M}}_t$. The map state therefore satisfies the semantic criterion of Section~\ref{sec:system}, as it admits a decoder to the referent $\hat{\boldsymbol{M}}_t$ under a distortion measure. The task state $\boldsymbol{\tau}_t^i$ has no such referent and is therefore task-oriented in the narrower sense. In case no message is received, we let $\boldsymbol{m}_t^{\mu,i} = \boldsymbol{0}$, ensuring $\boldsymbol{\mu}_{t}^i$ is never updated with hallucinated information.

For the remaining updates, $\bar{\boldsymbol{m}}_t^i$ equals $\boldsymbol{m}_t^i$ when a message is received and a predicted substitute otherwise (Section~\ref{sec:pred}). In both cases, $\bar{\boldsymbol{m}}_t^i$ represents a complete aggregated message, with $\bar{\boldsymbol{m}}_t^{\mu,i}$ serving as its map component (Section~\ref{sec:comms}).

\subsubsection{Task state update} The task state is updated at each time step by a recurrent function $\boldsymbol{F}_{\boldsymbol{\theta}}$ that integrates the previous task and map states with the UAV's current local observation, proprioceptive state, and the best available aggregated message, $
    \boldsymbol{\tau}_t^i = \boldsymbol{F}_{\boldsymbol{\theta}}\!\left(\boldsymbol{\tau}_{t-1}^i,\;
    \boldsymbol{o}_t^i,\; \boldsymbol{e}_t^i,\; \boldsymbol{\mu}_{t-1}^i,\;
    \bar{\boldsymbol{m}}_t^{i}\right)$.

\subsubsection{Memory state update} 

The memory state is updated at each time step by a recurrent model $\boldsymbol{G}_{\boldsymbol{\psi}}$ applied to the previous memory and task states, local observation, and best available aggregated message from time $t-1$, 
$\boldsymbol{\rho}^i_t = \boldsymbol{G}_{\boldsymbol{\psi}}\!\left(\boldsymbol{\rho}^i_{t-1},\, \boldsymbol{\tau}^i_{t-1},\, \boldsymbol{o}^i_{t-1},\, \bar{\boldsymbol{m}}^i_{t-1}\right)$. By conditioning on $\bar{\boldsymbol{m}}^i_{t-1}$, the memory state models the dynamics of swarm information flow beyond what local observations reveal. 


In our implementation, $\boldsymbol{G}_{\boldsymbol{\psi}}$, as well as $\boldsymbol{E}_{\boldsymbol{\phi}}$ and $\boldsymbol{F}_{\boldsymbol{\theta}}$, will be realized as gated recurrent units~\cite{cho2014gru}.

\subsection{Message Aggregation}
\label{sec:comms}

Each UAV $j$ broadcasts the message $\boldsymbol{z}_{t-1}^j = \left[\boldsymbol{\mu}_{t-1}^j; \boldsymbol{\tau}_{t-1}^j\right]$ at each time step. Since state update functions require fixed-size inputs, the set $\mathcal{M}_t^i = \left\{\boldsymbol{z}_{t-1}^j : j \in \mathcal{N}_t^i\right\}$ must be reduced to a permutation-invariant summary vector $\boldsymbol{m}_t^i \in \mathbb{R}^{d}$. We adopt mean pooling as the aggregation mechanism,
$\boldsymbol{m}_t^i = \frac{1}{|\mathcal{N}_t^i|} \sum_{j \in \mathcal{N}_t^i}\boldsymbol{z}_{t-1}^j$. Mean pooling is parameter-free, permutation-invariant, and preserves the structure of the aggregated message. 
The resulting vector $\boldsymbol{m}_t^i = [\boldsymbol{m}_t^{\mu,i},\; \boldsymbol{m}_t^{\tau,i}]$ decomposes naturally into map and task components. Only the map component $\boldsymbol{m}_t^{\mu,i}$ is passed to $\boldsymbol{E}_{\boldsymbol{\phi}}$, while the full message $\boldsymbol{m}_t^i$ is used by the remaining state updates. 



\subsection{Memory-based Message Prediction}
\label{sec:pred}

When UAV $i$ loses contact with all neighbors (i.e., $\mathcal{N}_t^i = \emptyset$), a generative model $\boldsymbol{P}_{\boldsymbol{\zeta}}$ 
is used to predict the aggregated message the UAV expects to receive. Since $\boldsymbol{\rho}_t^i$ accumulates information only up to time $t-1$, the latest local observation $\boldsymbol{o}_t^i$ and proprioceptive state $\boldsymbol{e}_t^i$ are additionally provided to incorporate the most current sensory information via
$\hat{\boldsymbol{m}}_t^i = \boldsymbol{P}_{\boldsymbol{\zeta}}\!\left(\boldsymbol{\rho}_t^i,\;
    \boldsymbol{o}_t^i,\; \boldsymbol{e}_t^i\right)$.
Thus, the best available message $\bar{\boldsymbol{m}}_t^i$ equals $\boldsymbol{m}_t^i$ when $\mathcal{N}_t^i \neq \emptyset$, or $\hat{\boldsymbol{m}}_t^i$ otherwise, ensuring that decisions are taken on the most up-to-date swarm representation. 

\subsection{Actor-Critic Architecture}
\label{sec:policy}

Each UAV selects its action through an actor-policy $\pi_{\boldsymbol{\xi}}(a^i_t \mid \tau^i_t)$ that maps its current task state $\tau^i_t$ to a distribution over $\mathcal{A}^i$. Since $\tau^i_t$ encodes the UAV's local observations, spatial awareness, and best available message, the policy operates identically whether or not communication was available at time $t$. As all UAVs are homogeneous in hardware and task structure, the policy parameters $\boldsymbol{\xi}$ are shared across the swarm. Model sharing improves sample efficiency and scales to swarms of arbitrary size. In our implementation, the actor network maps $\tau^i_t$ to the mean and diagonal covariance of a Gaussian distribution, from which velocity and heading commands are sampled at execution time. Training $\pi_{\boldsymbol{\xi}}$ on the shared reward $r_t$ alone yields noisy gradients, since $r_t$ depends on all UAVs' joint actions~\cite{sutton2018}. Thus, we employ the \textit{actor-critic} framework with a centralized critic $V_{\boldsymbol{\omega}}(\boldsymbol{s}_t)$. The actor is updated with the advantage $\hat{A}_t = R_t - V_{\boldsymbol{\omega}}(\boldsymbol{s}_t)$, where $R_t = \sum_{k=t}^{T} \gamma^{k-t} r_k$, providing low-variance estimates for stabilized training~\cite{konda2000, lowe2017maddpg}. We adopt the multi-agent proximal policy optimization (MAPPO)~\cite{mappo} algorithm, as its on-policy nature is compatible with our recurrent updates. 

\subsection{Joint Training Objective}
\label{sec:training}

We propose training the entire architecture end-to-end centrally, combining MAPPO's objective with self-supervised losses for map reconstruction and message prediction
\begin{equation}
\begin{split}
    \small
    \mathcal{L}(\boldsymbol{\xi}, \boldsymbol{\omega}, \boldsymbol{\phi},
    \boldsymbol{\eta}, \boldsymbol{\theta}, \boldsymbol{\psi}, \boldsymbol{\zeta}) =
    -\mathcal{L}_{\text{CLIP}}(\boldsymbol{\xi}) +
    &c_1\, \mathcal{L}_{\text{VF}}(\boldsymbol{\omega})\\
    + c_2\,& \mathcal{L}_{\text{map}}(\boldsymbol{\phi}, \boldsymbol{\eta}) +
    c_3\, \mathcal{L}_{\text{pred}}(\boldsymbol{\zeta}, \boldsymbol{\psi}),
    \end{split}
    \raisetag{9.8mm}
\end{equation}
where $\mathcal{L}_{\text{CLIP}}(\boldsymbol{\xi})$ is the clipped surrogate objective \cite{mappo}, $\mathcal{L}_{\text{VF}}(\boldsymbol{\omega}) = \frac{1}{T}\sum_{t=1}^T \left(R_t - V_{\boldsymbol{\omega}}(\boldsymbol{s}_t)\right)^2$ is the critic loss, and $c_1, c_2, c_3 > 0$ are scalar hyperparameters that balance the relative importance of each objective. Note that $\boldsymbol{\theta}$ is updated implicitly through backpropagation.

The map loss trains $\boldsymbol{E}_{\boldsymbol{\phi}}$ and $\boldsymbol{D}_{\boldsymbol{\eta}}$ to reconstruct $\hat{\boldsymbol{M}}_t$ via 
\begin{equation}
\small
    \mathcal{L}_{\text{map}}(\boldsymbol{\phi}, \boldsymbol{\eta}) =
    \frac{1}{NT}\sum_{i=1}^N\sum_{t=1}^T
    \left\|\boldsymbol{D}_{\boldsymbol{\eta}}\!\left(\boldsymbol{\mu}_t^i\right) -
    \hat{\boldsymbol{M}}_t\right\|^2_F.
\end{equation}
This ensures that each UAV's map state $\boldsymbol{\mu}_t^i$ serves as a reliable compressed representation of the current global map estimate, as required by the map state update in Section~\ref{sec:latent}. The prediction loss measures the mean squared error between received aggregated messages and their predictions, over time
\begin{equation}
\small
    \mathcal{L}_{\text{pred}}(\boldsymbol{\zeta}, \boldsymbol{\psi}) = \frac{1}{N} \sum_{i \in \mathcal{N}}
    \frac{1}{|\mathcal{T}_{\text{comm}}^i|} \sum_{t \in \mathcal{T}_{\text{comm}}^i}
    \left\|\boldsymbol{m}_t^i - \hat{\boldsymbol{m}}_t^i\right\|^2,
\end{equation}
where $\mathcal{T}_{\text{comm}}^i = \{t : \mathcal{N}_t^i \neq \emptyset\}$ is the set of time steps at which UAV $i$ has at least one active neighbor. Training follows a standard on-policy rollout procedure, with dropouts simulated by sampling link states per Section~\ref{sec:comm}. The joint loss $\mathcal{L}$ is computed over each batch and a single gradient update is applied to all parameters simultaneously, until convergence.


\section{Simulation Results and Analysis}
\label{sec:experiments}

We evaluate our framework in a $50 \times 50$ discrete grid environment with a swarm of $N = 3$ UAVs, each equipped with a sensor range of 2 cells. UAVs are initialized at random positions, at an altitude of 10\,m with zero initial velocity and uniformly random headings, and they execute continuous velocity commands over episodes of $T = 1000$ steps ($\kappa = 0.1$). During motion, UAV altitude is constrained to $[5, 20]$\,m and horizontal positions are clipped to the arena bounds. The ground-truth occupancy map is generated independently for each episode by placing 30 occupied cells uniformly at random over the grid. The A2A channel follows the model of Section~\ref{sec:comm}, with Rician factor $K = 5$, $L^{\text{atm}} = 0.9$, {$P^j = 20$\,dBm}, and $\alpha = 2$. We set the SINR detection threshold to $\beta_{\mathrm{th}} = 0.05$ and the noise power to $\sigma^2 = 1$, and draw the interference as $I \sim \mathcal{U}(0, 0.5)$. A link failure probability of $p_f = 0.3$ is used during training. As this paper presents a proof of concept, these values are chosen to pose a meaningful challenge to the framework without pushing it to its limits. All latent states have dimension $d_\mu = d_\tau = d_\rho = 512$. The framework is trained for 1600 epochs of 600 steps each across 40 parallel environments using MAPPO, with discount factor $\gamma = 0.99$, GAE parameter $\lambda = 0.95$, clipping parameter $\epsilon = 0.2$, and entropy coefficient $0.01$. The reward weights are $w_{cov}=0.4, w_{red}=0.02, w_{coll}=0.8$, while the loss weights are set to $c_1=0.5$ and $c_2 = c_3 = 5$. The learning rate is $10^{-4}$ for the first 1000 epochs and $10^{-5}$ thereafter. All results are averaged over 10 Monte Carlo runs.

\begin{table*}[ht!]
\centering
\caption{Ablation study results (left) and robustness to communication dropout (right), averaged over 10 Monte Carlo runs.\vspace{-2.5mm}}
\label{tab:results}
\begin{minipage}{0.5\textwidth}
\centering
\resizebox{\linewidth}{!}{%
\begin{tabular}{lccccc}
\toprule
\textbf{Condition} & \textbf{Coverage (\%)} & \textbf{Collisions}  & \textbf{Steps to 90\%} \\
\midrule
Full comm (UB)      & $92.40 \pm 2.64$ & $10.30 \pm 12.54$ & $732.71 \pm 90.58$ \\
\textbf{Proposed (full)} & $\mathbf{90.98 \pm 2.92}$ & $\mathbf{8.70 \pm 8.49}$ & $\mathbf{692.71 \pm 189.04}$ \\
No memory           & $81.90 \pm 5.51$ & $41.50 \pm 24.08$ & N/A \\
No comm (LB)        & $83.44 \pm 4.80$ & $31.90 \pm 21.85$ & N/A \\
No predictor        & $70.30 \pm 3.61$ & $54.90 \pm 26.12$ & N/A \\
\bottomrule
\end{tabular}}
\end{minipage}
\hfill
\begin{minipage}{0.28\textwidth}
\centering
\resizebox{\linewidth}{!}{%
\begin{tabular}{ccc}
\toprule
$p_f$ & \textbf{Coverage (\%)} & \textbf{Steps to 90\%} \\
\midrule
0.00 & $91.66 \pm 2.33$ & 665 \\
0.05 & $92.59 \pm 2.04$ & 593 \\
0.15 & $91.36 \pm 2.00$ & 758 \\
0.30 & $90.89 \pm 2.18$ & 651 \\
0.50 & $90.96 \pm 2.17$ & 677 \\
\bottomrule
\end{tabular}}
\end{minipage}\hspace{16mm}\vspace{-0mm}
\label{tab:ablation}
\end{table*}


To assess each proposed component's contribution, we compare five conditions. \textit{Proposed (full)} corresponds to the complete framework at test time. \textit{No predictor} replaces $\hat{\boldsymbol{m}}^i_t$ with a zero vector during dropout. \textit{No memory} zeroes $\boldsymbol{\rho}^i_t$ before it gets used in any update. \textit{Full comm} forces all links active at every step, serving as a coordination upper bound. \textit{No comm} forces all links inactive throughout, serving as a lower bound. 

Table~\ref{tab:results} presents the final coverage, collisions, and steps to 90\% coverage for the five conditions. The proposed model achieves $90.98 \pm 2.92\%$ final coverage, within $1.5\%$ of the full communication upper bound. These results show that latent-space prediction alone nearly recovers uninterrupted communication performance. Removing the message predictor is the most harmful ablation, reducing coverage to $70.30\pm3.61\%$ and raising collisions nearly sixfold ($8.70$ vs $54.90$). The lack of a predictor deteriorates performance more than No comm ($83.44\%$, $31.90$ collisions), which shows that an all-zero vector in place of the missing message actively misleads the swarm. Removing the memory is less severe but still substantial ($81.90\pm5.51\%$). Without the accumulated history, the predictor loses the context it conditions on, and the resulting degradation indicates that prediction quality depends on memory rather than current observations alone. Neither ablation reaches $90\%$ coverage within the episode horizon, whereas the proposed model matches the full communication upper bound in collisions ($8.70$ vs $10.30$). Table~\ref{tab:ablation} also presents coverage under a sweep of $p_f \in \{0.00, 0.05, 0.15, 0.30, 0.50\}$. Our approach maintains consistently high coverage ranging from $90.89\%$ to $92.59\%$, with low variance and no monotonic degradation with $p_f$, showing that the latent-space predictor sustains coordination even under severe link failures.

\vspace{-1mm}

\section{Conclusion}
\label{sec:conclusion}

In this paper, we have proposed a memory-augmented framework for cooperative UAV reconnaissance under intermittent A2A communication. Each UAV maintains a structured latent state decomposed into map, task, and memory components. During dropout, a generative predictor infers a substitute for the aggregated peer message from the UAV's accumulated history in the latent space. An asymmetric strategy shields map estimates from hallucinated updates. Simulation results show that the proposed approach achieves $90.98\%$ final coverage under $p_f = 0.3$, closing the gap to the full communication upper bound to less than $1.5\%$. Performance remains stable across link failure probabilities up to $50\%$. Our ablations show both the memory state and the predictor to be essential, as removing either degrades coverage and increases collisions. Future work includes benchmarking against explicit-communication mapping approaches \cite{mahdoui2020, westheider2023}, and extending our approach to larger swarms and heterogeneous platforms.

\vspace{-1mm}

\end{document}